# Testing the ability of Multivariate Hybrid Spatial Network Analysis to predict the effect of a major urban redevelopment on pedestrian flows

Crispin H. V. Cooper, Sustainable Places Research Institute, Cardiff University, 33 Park Place, Cardiff CF10 3BA cooperch@cardiff.ac.uk

Ian Harvey, Data Innovation Research Institute, Cardiff University, Trevithick Building, The Parade, Cardiff, CF24 3AA; WISERD, Cardiff University, 38 Park Place, Cardiff, CF10 3BB
harveyic@cardiff.ac.uk

Scott Orford, School of Geography and Planning, Cardiff University, Glamorgan Building, King Edward VII Avenue, Cardiff, CF10 3WA; WISERD, Cardiff University, 38 Park Place, Cardiff, CF10 3BB
orfords@cardiff.ac.uk

Alain J. Chiaradia, Faculty of Architecture, University of Hong Kong, 4/F, Knowles Building, The University of Hong Kong, Pokfulam Road, Hong Kong alainjfc@hku.hk

## Abstract
Predicting how changes to the urban environment will affect town centre vitality, mediated as pedestrian flows, is important for environmental, social and economic sustainability. This study is a longitudinal investigation of before and after urban environmental change in a town centre and its association with vitality. The case study baseline is Cardiff town centre in 2007, prior and after major changes instigated by re-configuring Cardiff public and quasi-public street layout due to implementation of the St David's Phase 2 retail development.

We present a Multivariate Hybrid Spatial Network Analysis (MHSpNA) model, which bridges the gap between existing Spatial Network Analysis models and four stage modelling techniques. Multiple theoretical flows are computed based on retail floor area (everywhere to shops, shop to shop, stations to shops and parking to shops). The calibration process determines a suitable balance of these to best match observed pedestrian flows, using generalized cross-validation to prevent overfit. Validation shows that the 2007 model successfully predicts vitality as pedestrian flows measured in 2010 and 2011. This is the first time, to our knowledge, that a vitality-pedestrian flow model has been evaluated for its ability to forecast town centre vitality changes over time.

## Introduction
Predicting how changes to the urban environment will affect vitality mediated as pedestrian flows is important for numerous reasons. From a sustainable transport perspective, substitution of motorized trips with walking is not only beneficial for our ecological and carbon footprint (Frank and Pivo 1994; Cervero and Kockelman 1997), but also reduces congestion and air pollution, increases community cohesion (Cooper, Fone, and Chiaradia 2014), and - in the face of an obesity crisis - improves public health (Handy et al. 2002; Handy 2005).

Pedestrian footfall is also key to understanding town centre vitality and hence economic sustainability. UK policy (Department for Communities and Local Government 2009) stresses the importance of 'linked trips'. The progressive tightening of retail planning regulation in the decade that followed the National Planning Policy Framework (2012), and retailer adaptation to that tightening, resulted in 'town centre first' approaches to retail planning policy. Since then academic research has started to suggest a more positive role for such developments than hitherto, and to indicate that they can play an important role in anchoring small centres (Lambiri, Faggian, and Wrigley 2017). One of the key aims is to ensure that town centre strategy provides a vision to enhance vitality and viability. Vitality is usually measured by pedestrian flows (Department for Communities and Local Government 2014 paragraph 5). To this end, town centres have audited pedestrian volumes and their changes over time.

Available evidence on the impact of the 'town centre first' approach on vitality is scant. Our purpose in this paper is to begin the process of presenting that missing evidence. In particular, for the first time we analyse the vitality - pedestrian flow data before and after the implementation of a large re-configuration of street layout of the centre of Cardiff, the capital of Wales. The St David's Phase 2 development enclosed several new arcades and re-routed existing streets. The re-organisation of the street layout decreased urban block size, making it more permeable, with dual indoor and outdoor frontage and clarified the structure of the town centre. Survey data shows that the outdoor pedestrian level remained similar or increased but does not show how pedestrian flows and hence vitality are redistributed in the new layout. Unlike previous studies (Sung, Go, and Choi 2013; Sung and Lee 2015) in this paper we contribute the first longitudinal vitality - pedestrian flow model able to predict the effect of changes in layout on pedestrian flow distribution.

Few published vitality-pedestrian models include an empirical test against observed pedestrian flow data. Those which do, invariably test model fit at a single point in time as a cross-sectional study. Longitudinal studies, which track changes over time, are considered more reliable (Greenhalgh 1997) but have not been used in this field. The recording of pedestrian flows before and after a major change to the urban environment brings the opportunity to raise the bar: given a model calibrated on data prior to the change, can the effects of the change be predicted correctly? We thus test the vitality model's ability to extrapolate across time. To our knowledge this is the first time any vitality-pedestrian flow model in the above category (3) has been tested in this way. We also, using generalized cross-validation, test our ability to extrapolate across space.

We describe our own model as Multivariate Hybrid Spatial Network Analysis (MHSpNA). It aims to fill a gap between existing SpNA models, and traditional transport four stage models that sequentially determine origin and determination of trips, journey distribution matching origins with destinations, mode choice, route assignment. This could at first glance be characterised as a continuum between sketch planning (less accurate/less expensive models using sparse data suitable for early stage urban design projects), and more accurate/more expensive models suited to major transportation infrastructure and detailed design phases. However, the more expensive option is not without its own flaws. Transferring methods geared around predicting vehicular flow has historically required exclusion of minor streets from the representation of the built environment, and aggregation of trips into large analysis zones, neither of which is acceptable for analysis of pedestrians whose trips may fall entirely within a single zone (Cervero 2006). Exclusion of land-use/accessibility feedback cycles has led to historic failures in vehicle modelling (Atkins 2006 dissects one such example) and this cycle is known to be relevant to pedestrian models through residential self-selection (Cervero 2006). These problems are not insoluble, but require substantial extra cost in data collection, calibration and computation, and research in this area is ongoing (Department for Transport 2014b section 4.6.6, see also 2014c). SpNA meanwhile utilizes all the street network available to pedestrians, does not require zones, and is thought to capture land use feedbacks at least approximately (Chiaradia, Cooper, and Wedderburn 2014; Cooper 2017). Our SpNA model is calibrated directly against observed flows, meaning data is not needed to calibrate separate submodels of trip generation and distribution, mode choice and assignment; however verification of these is likely to lead to a more accurate model if data is available. It is also a unimodal model; while Department for Transport (2014a) allows taking this approach for simplicity, in principle, extension to a multimodal network is also possible.

Alternatively, MHSpNA as used here can be characterised as a heterogeneous agent model, as it simulates the journeys of numerous agents through the network with differing goals and preferences. We note that some transport practitioners prefer to reserve the term 'agent model' for models of individually interacting agents which MHSpNA is not. However this lack of inter-agent interaction carries the advantage that multiple models may be more easily calibrated by linear regression.

## Case Study and Data

The case study area is Cardiff, the capital and largest city in Wales, UK. Cardiff has a population of around 350 thousand, or around 10% of the Welsh population. In terms of urban morphology and demographic characteristics it is typical of a medium size British city. However, its capital city status means that it is a major retail and tourist destination and over the past 10 years the city centre has undergone a major regeneration programme – see Figure 3. A new shopping mall, St David's Centre opened in 2009, and the surrounding shopping streets were pedestrianised. The city Centre also includes Victorian and Edwardian shopping arcades and a civic centre to the north which hosts the University, Law Courts and Government buildings. To the south of the city centre is the main bus station and Central railway station and there are a number of car parks to the north, east and south. Bordering the city centre to the south west is the national stadium for Wales, the Millennium Stadium, which opened in 1999 and to the north west is Cardiff Castle and Bute Park.

Collection of pedestrian flow data was commissioned by Cardiff Council for most years between 1999 and 2011. This effort was commissioned for the purpose of examining retail footfall and the town centre vitality profile, meaning that pedestrian counting is limited to medium- to high-flow streets. Ideally to calibrate a vitality-footfall model, a stratified sample is required which also includes low-flow streets. By necessity we make do without such data. We calibrate the pedestrian model to summer 2007, the last snapshot available before work on the St David's Centre began in winter 2007. We test the model on the years 2010 and 2011, the only years available after completion of the development. Each year comprises data collected on Thursday evening, Friday, Saturday and Sunday daytimes (10am-4pm i.e. not including peak commuting times). As we focus on vitality-shopping patterns, we model combined Friday and Saturday flows, excluding Thursday evening and Sunday as not being representative of typical optional town centre visit behaviour. The trend over time of this data is shown in Figure 1. We take the year-on-year change in the mean to be caused by exogenous factors such as the 2008 financial crash and subsequent recovery, and therefore do not try to model changes in the mean. Thus we report success of model predictions as correlation ($r^2$) which discards scale information, rather than mean square error which would preserve scale but principally be measuring factors external to the model.

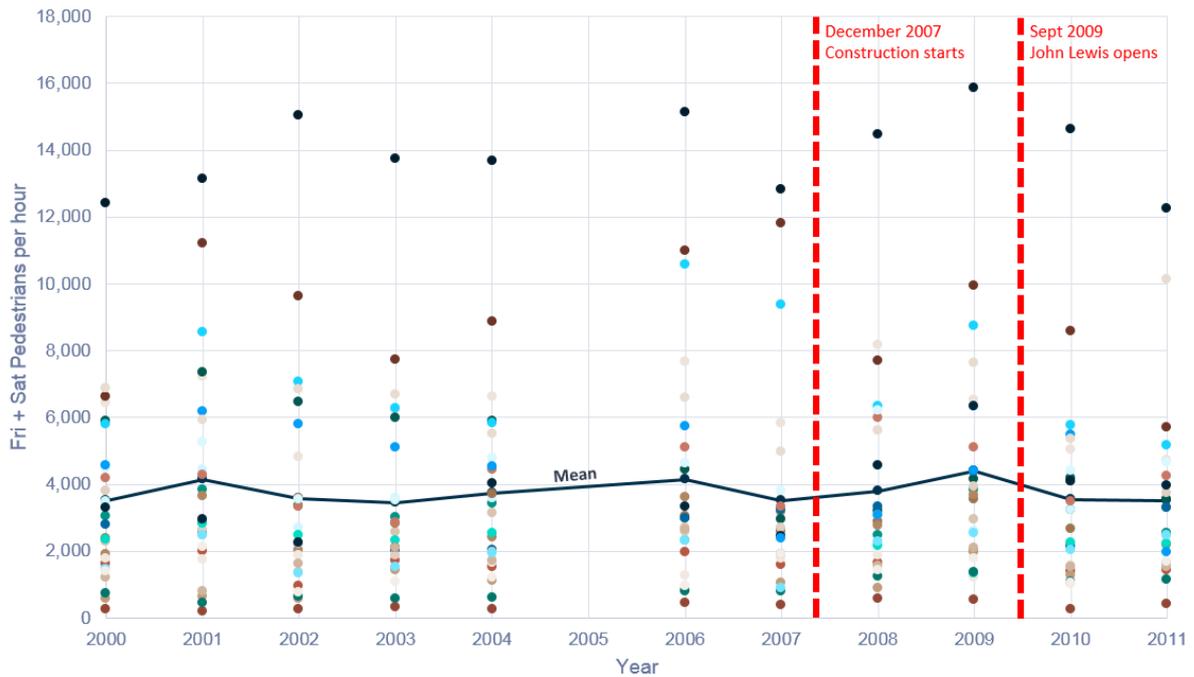

*Figure 1 Cardiff City Centre Pedestrian Survey Data (only showing points repeated across all years)*

A pedestrian network is a topological map that contains the geometric relationship between pedestrian path segments (e.g., sidewalk, crosswalk, and footpath), which is needed in a variety of applications such as pedestrian navigation systems/services, urban planning (Karimi and Kasemsuppakorn 2013) and pedestrian urban design. Different approaches to develop pedestrian network maps have been attempted (Elias 2007; Karimi and Kasemsuppakorn 2013). Given the reported effort, cost and complexity of automated generation in non-regular street patterns (Karimi and Kasemsuppakorn 2013) manual network generation with field surveys appears to be a popular approach (Chiaradia 2013). This method has been used in this paper with ArcGIS tools. A pedestrian path segment mapped as a link is any pathway between two junctions that allows pedestrians to pass and can be categorized into types such as: sidewalk, crosswalk, footpath, building public path (arcade, shopping mall main path), trail, pedestrian bridge, and tunnel. The vector data model, due to its ability to represent complex spatial objects using basic graphical elements (points and lines), is suitable for representing pedestrian networks. Pedestrian network mapping is a generic mapping process like the process of generating a Linear Referencing System (Curtin, Nicoara, and Arifin 2007).

The authors used the Historic Mastermap (Ordnance Survey 2017) in ArcGIS to re-draw the outdoor and indoor pedestrian network for 2007 and 2011. The network extends into the surrounding area via a 1.2km buffer to serve as source for trips from the surroundings (classed as 'everywhere' under the definition of variables in Table 1). Using the background vector map and available floor plans, links and nodes of the pedestrian network were constructed by manual drawing. The following assumptions guided the pedestrian network mapping:

1. Links were drawn with the assumption that path centreline is representative of pedestrian thoroughfare.
2. Gradients in paths were ignored as Cardiff town centre has very low gradient and can be considered as mainly flat terrain.

3. Given assumption 1, field surveys and publicly indoor displayed floor plans provide a fair indication of the real features of indoor pedestrian paths that function as quasi-public paths, such as traditional and new shopping arcades and malls.
4. In streets which do not have formalised crossings, such as in residential areas, the pedestrian network is mapped as road centre line and junction. In residential areas, streets without crossings receive low traffic, hence for pedestrians it is easy to cross from one sidewalk to the opposite sidewalk.
5. The street mapped with road centre lines (from 4. above) when linking with a street with pedestrian crossing are mapped as in Figure 2. The crossing is itself considered as a link because it is an area of interaction with vehicular traffic.

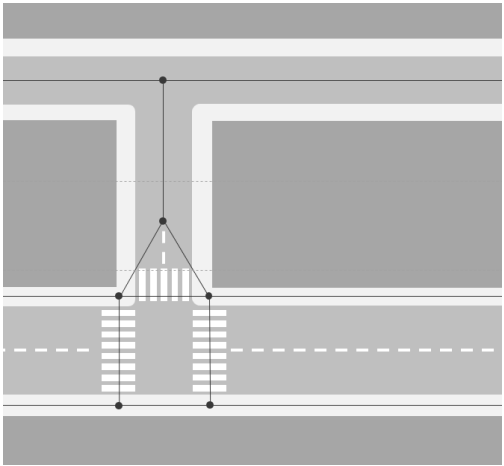

*Figure 2 Principle of pedestrian network mapping – link and node: residential streets with low traffic and no formalised crossing linked to a street with formalised crossings.*

A link has the following attributes: length, angular curvature change along the link. The purpose is to include key attributes to pedestrian route choice such as Euclidean distance and directness that are key to pedestrian navigation (Montello 1998, 2005) in open large scale urban environment. Route choice is one of the processes of pedestrian navigation that may be described by general choice theory. A route is described as a chain of consecutive nodes, the junctions, joined by links, connecting trip origin and destination (Bovy and Stern 1990). A wide variety of algorithms have been developed to represent the route choice decision-making processes for different transport modes (highways, public transport, cycling, walking). Although route selection strategies are largely subconscious (Hill 1984), several researchers have formulated theories on this behaviour. Distance is not only an important factor on which route choice is based, it also influences the way pedestrians choose their routes in pedestrian areas (Ciolek 1978; Hill 1982). Verlander and Heydecker (1997) showed that 75% of pedestrians were taking the shortest path, but did not check whether the shortest Euclidean paths were also the most direct paths which is often the case (Zhang, Chiaradia, and Zhuang 2015). Different types of distance are therefore distinguished in the literature. Khisty (1999) distinguishes between 'perceived' distance and 'cognitive' distance which include the assessment of the geometric complexity of the routes (directness). This is linked with the concept of visibility in that pedestrians tend to walk straight towards a visible destination, unless they are hindered by obstacles, other pedestrians, or diverted by other attractions; however we do not measure visibility directly. In city centres, Lausto & Murole (1974) have shown the importance of retail and public transport service points; hence two nearby railway stations (Cardiff Central and Queen Street) were added to the network along with estimates of retail floor area.

Retail floor area was derived from business rates data. Business rates is the commonly used name in England and Wales of non-domestic rates, a tax on the occupation of non-domestic property. Rates information is held by the Valuation Office Agency (VOA), and data is queryable online via the tax.service.gov.uk website for the period 2010-2016. Information available includes the full address of the property, a free-text description of use of the property, the total taxable area (m² / unit), the price per area (m² / unit) and the current rateable value. Properties were extracted from the website for the Cardiff local authority area. Postcodes of properties were used in the Google Geocoding API address lookup service to retrieve an address point; where this was unavailable, the OS Open data postcode midpoint was used. The properties for Cardiff city centre were then extracted using the point data. The description of the use of the property in the dataset was used to identify retail and leisure outlets. String "fuzzy matching", utilising the Levenshtein (1965) distance algorithm, was used to discover similar description of strings for grouping into retail types e.g. "pubs, public houses, nightclubs". The floor area for each address point was added to the network. Carparks exceeding capacity of 500 spaces were also identified from the business rates data and added to the model.

The Business Rates data did not contain information on buildings that had been demolished before 2010, which included the re-developed area. Here building layout and floor area were reconstructed via other data sources including historic OS Mastermap, historic Google Earth aerial photographs and local knowledge. This allowed a before (baseline) and after (future condition) model of the St David's 2 development to be constructed.

Figure 3 maps the data available to the model.

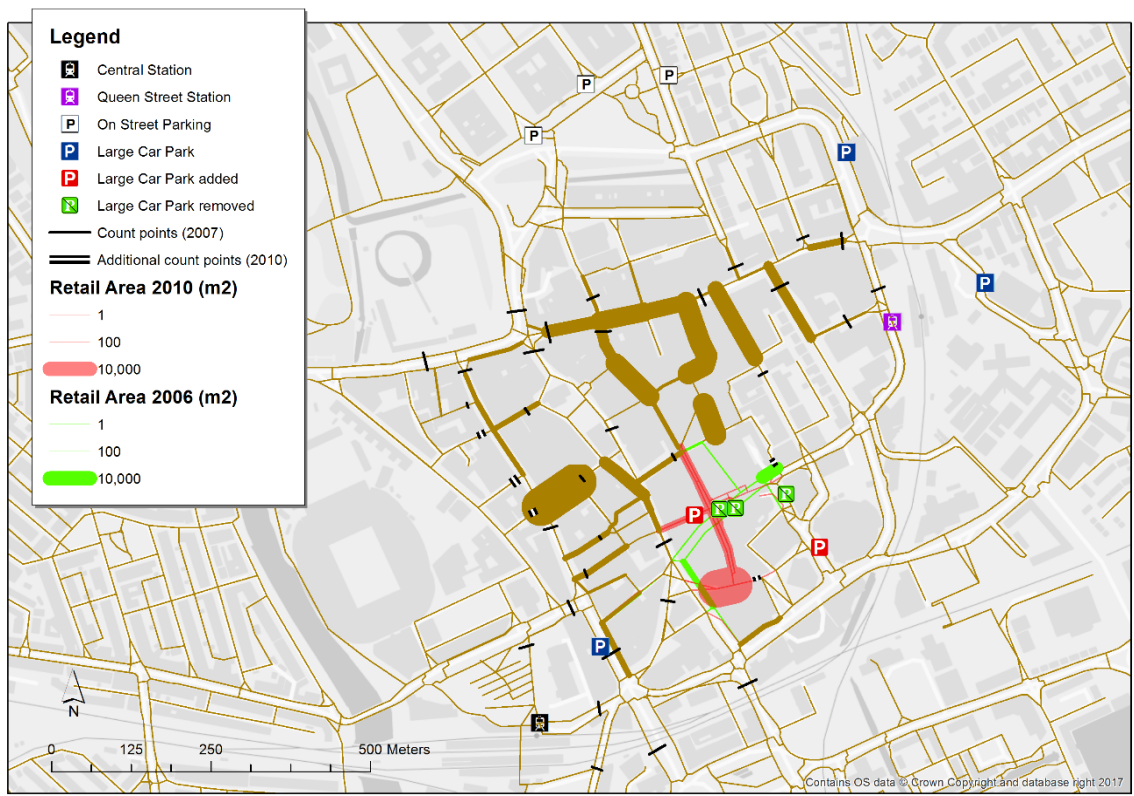

*Figure 3 Summary of changes to network, retail area, car parks and pedestrian flow measurement points 2007-2010*

## Methods

### Multiple Hybrid Betweenness

The building block for the MHSpNA model is the SpNA measure of Betweenness. In simple terms, this is a flow model derived by summing 'shortest' paths from 'everywhere to everywhere' subject to a constraint on maximum pedestrian journey distance. It can also be characterized as a direct demand transport model in which journey generation, distribution and mode choice are considered congruent (Cervero 2006; Lowry 2014; Ortúzar and Willumsen 2011, chapter 6).

Note that the above definitions of 'shortest' and 'everywhere' may vary. We modify our definition of 'shortest' to incorporate distance that is neither purely angular nor Euclidean, as with SpNA tradition (Hillier and Iida 2005), and instead use a hybrid of both distance types also including a random component. We modify 'everywhere' to specific sets of origins and destinations, but repeat this process for multiple journey types (shown below in Table 1), hence making the model multivariate.

The formula for Betweenness used by the sDNA software assumes uniqueness of shortest paths and is given in Equation 1:

$$Betweenness\ (x, rmin, rmax, d_{routing}, d_{radius}) = \sum_{y \in O} \sum_{z \in D \cap R(y, rmin, rmax, d_{radius})} W(y,z) OD(y,z,x,d_{routing}) \quad (1)$$

where x, y and z are links in the network N, O is the set of links defined as origins, D is the set of links defined as destinations, and W(y,z) is the weighting of a trip from y to z. $R(y, rmin, rmax, d_{radius})$ is the subset of the network closer to link y than a threshold radius *rmax* but further from y than *rmin* according to $d_{radius}$. $OD(y,z,x,d_{routing})$ is defined in Equation 2:

$$OD(y,z,x,d_{routing}) = \begin{cases} 1 & \text{if } x \text{ is on the shortest path from } y \text{ to } z \text{ as defined by metric } d_{routing} \\ 1/2 & \text{if } x = y \neq z \text{ or } x = z \neq y \quad \text{(partial contribution for endpoints)} \\ 1/3 & \text{if } x = y = z \quad \text{(partial contribution for self} - betweenness) \\ 0 & \text{otherwise} \end{cases} \quad (2)$$

$d_{routing}$ and $d_{radius}$ are metrics defining what we mean by 'distance'. $d_{radius}$ is consulted when deciding whether a journey of a certain distance takes places at all (in the current computation) and for the current study is defined as Network Euclidean distance i.e. the shortest possible distance measured along the network in metres. $d_{routing}$ is consulted to determine which route the journey will take; the definition is different and is given in the next section. Note that the different definitions of $d_{routing}$ and $d_{radius}$ mean that occasionally the routes taken by journeys will be longer than the distance band they are supposed to represent. This seeming inconsistency does not cause problems in practice (Cooper 2015) and is chosen because the simple definition of $d_{radius}$ as reach makes results easier to interpret.

The usual approach to defining the journey weighting function is to set W(y,z)=W(y)W(z) where W(y) and W(z) are the weights of the origin and destination respectively (as defined in Table 1). Assuming uniform distribution of origins, destinations and weights across space this implies that total journey activity scales with the square of the average total weight within each radius (Equation 3):

$$total\ trip\ activity = \sum_{y \in O}(W(y) \sum_{z \in D \cap R} W(z)) \approx \left( \frac{\sum_{y \in O}(W(y) \sum_{z \in D \cap R} W(z))}{\sum_{y \in O}(W(y))} \right)^2 \quad (3)$$

It is therefore a fully elastic measure of demand, at least with respect to distribution of opportunity across space, in that more opportunities for interaction generate more interaction. Depending on how the analysis is weighted, the unit of opportunity can either be a defined land use type such as a shop, or alternatively represent the network itself: in both cases the implicit assumption is that interaction is increased by intensification of activity from a given land use (e.g. greater attractivity and hence vitality per square metre of retail floor area); however the latter case can also be taken to include intensification of land use itself (e.g. more floor area, possibly on multiple levels).

In the current study we also introduce 'Two Phase Betweenness'. As the name suggests this is computed in two phases; (1) determine total accessible destinations; (2) distribute origin weight across available destinations. It is thus a fully inelastic model of demand (with respect to distribution of opportunity across space) in which trip volume is limited by the weight of the origin. Combining this feature with a limited radius (as we do) can also be interpreted as a form of intervening opportunity model. The formula is given in Equation 4, and implies that total trip activity scales with the average total weight in R rather than its square.

$$Two\ phase\ (inelaseic)\ W(y,z) = \frac{W(y)W(z)}{\sum_{z \in D \cap R} W(z)} \tag{4}$$

We also make use of Continuous Space Betweenness (Cooper and Chiaradia 2015) to improve accuracy for the smallest (200m) trip distances. Where betweenness flows stem from a single origin, there is no variety of opportunity across space, so the betweenness type is not relevant.

Table 1 shows the multiple types of betweenness combined to form the multivariate model. All analyses use the 'Polyline weighting' option in sDNA to interpret weights at face value rather than per unit length, link, etc.

| Key | Betweenness Type | Origin Weight | Destination Weight | Radii (max. trip lengths) |
|---|---|---|---|---|
| e2s | Elastic | Everywhere (network) | Retail Floor Area | 400, 800, 1200m |
| s2s | Inelastic, Continuous | Retail Floor Area | Retail Floor Area | 200, 400m |
| sq2s | Single Origin | Cardiff Queen Street Station | Retail Floor Area | 600, 1000m |
| sc2s | Single Origin | Cardiff Central Station | Retail Floor Area | 600, 1000m |
| p2s | Elastic | Large Car Parks | Retail Floor Area | 600, 1000m |
| n2s | Elastic | Dense area of on-street parking to North | Retail Floor Area | 600, 1000m |

*Table 1 Betweenness types incorporated in the model*

### Calibration of distance metric

The distance metric used for the study is given in Equation 5.

$$\begin{aligned} distance\ for\ link &= (a \times ang + (1-a) \times euc) \times rand \\ distance\ for\ junction &= a \times ang \times rand \end{aligned} \tag{5}$$

Where *ang* is cumulative angular change along a link or turn at junction; *euc* is Euclidean distance measured along the path of the link, and *rand* is a random sample drawn from a normal distribution with mean=1. The standard deviation of this distribution σ is varied to obtain optimal fit to pedestrian behaviour, the presumption being that typical pedestrian behaviour may not be random, but depends on more factors than we can feasibly include, so we randomize behaviour somewhat to ensure that pedestrian flows are distributed over similar paths rather than all-or-nothing assignment to the shortest. Thus for each origin-destination pair we draw multiple samples from the random distribution, 5 during the calibration of σ and 50 for the final model. To avoid distances of zero and the opposite extreme, values drawn from *rand* are constrained to the range 0.1 ≤ x ≤ 10; values exceeding that range are moved to its nearest endpoint.

For calibration of the random factor we test a wide range of values of σ for their effect on e2s Betweenness as a predictor of pedestrian flow (e2s being the variable which carries most predictive power on its own). The metric is also calibrated by varying the value of *a*, which specifies the hybrid between a purely angular metric (a=1) and purely Euclidean metric (a=0). From previous work we have found 0.25 ≤ a ≤ 0.5 to give good results, so tried both of these values (0.25, 0.5) during calibration. a=0.5 corresponds to the 'PEDESTRIAN' metric preset in sDNA and randomization was added using the *linerand, juncrand* and *oversample* keywords in sDNA advanced configuration.

### Calibration and testing of multiple models

The statistical process of model fitting is described in Appendix 1. Three models are formed and tested against each year 2010 and 2011:

1. The null model assumes no change in pedestrian flow between years, and is thus only applicable to points where pedestrian counts have been recorded all years.
2. The incremental model works by adding predicted change between years (derived from the regression model) to the flows for the baseline year. Like the null model, this cannot extrapolate from the count points to the rest of the network, where baseline flow data is not available. The incremental model is included to allow fair comparison of performance with the null model.
3. The direct model, as the name implies, uses regression outputs directly for prediction. This is the most useful model in practice as it extrapolates data across both space and time. Baseline flow data is only used for calibration and not as an input to the prediction of each flow point. It is therefore tested on 5 additional count points added by Cardiff Council for 2011, for which the null and incremental models could not have produced predictions at all.

In each case, the models are calibrated to the year 2007, and the coefficients derived applied to the changed map data for 2010 (adding/removing links, altering floor area and car park locations).

### Software

All modelling is undertaken with the publicly available sDNA+ toolbox for ArcGIS and QGIS (Cooper, Chiaradia, and Webster 2011; Cooper 2016). The *Prepare* tool is used to prepare the network and *Integral Analysis* for the betweenness models. Calibration via ridge regression is conducted using the open source sDNA *Learn* and *Predict* tools which in turn make use of the glmnet R package (Friedman, Hastie, and Tibshirani 2009).

### Results

During calibration to 2007 data we found setting the angular/Euclidean hybrid coefficient *a=0.5* to give better model fit than 0.25. For calibration of the random factor, Figure 4 shows the results obtained. Based on this we initially settled on σ=0.5 as the lowest amount of randomization that reliably increased correlation with pedestrian flows. However this led to predictions we considered unrealistic (such as all-or-nothing assignment to either side of a particular street with both were suitable), so we changed it to σ=1.0. This change also increased overall fit for the 2007 model from 0.47 to 0.49.

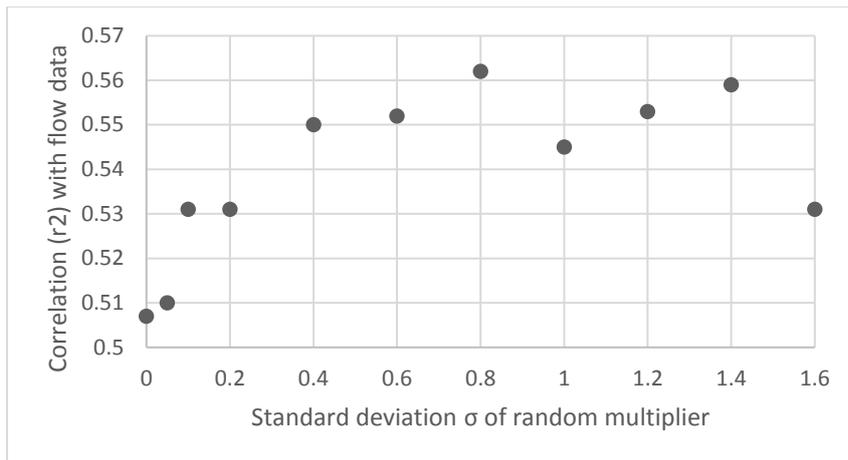

*Figure 4 Test of different levels of randomization on e2s variable, radius 800m, 5x sampling*

For the overall 2007 model, we initially fitted the first 5 variables described in Table 1. Inspection of the initial model fit revealed a number of errors in the map (connectivity and poor geocoding of retail floor areas), which were corrected. The final variable ('n2s') was added after the first calibration attempt, as we suspected from examination of residuals that the model was not capturing the large volume of on-street parking to the north of the study area, which is captured elsewhere through the 'e2s' variable. Additionally, two outliers at a single intersection which appears unusually busy, caused problems fitting the data, so the model was fit with the weighting λ=0.7 to reduce their effect (from previous work we have found values of λ in this region to improve GEH). The weighted, cross validated $r^2$ for the 2007 model is 0.49 including outliers; unweighted fit improves to 0.60 if removing outliers to test the outlier-fitted model. Standardized coefficients for the variables are given in Table 2. Ridge regression drops the p2s variable at 600m radius. Distance decay is evident in the coefficients (representing reduced tendency for individuals to travel further per opportunity), but not the standardized coefficients within the range of distances tested, as quantity of opportunities increases with distance.

| Variable | Radius (m) | Coeff | Std | StdCoeff |
|---|---|---|---|---|
| n2s | 600 | 4.0E-03 | 2.6E+04 | 102 |
| n2s | 1000 | 3.7E-03 | 7.4E+04 | 270 |
| p2s | 1000 | 1.8E-03 | 1.0E+05 | 187 |
| sc2s | 600 | 3.5E-03 | 1.7E+04 | 59 |
| sc2s | 1000 | 1.2E-03 | 4.8E+04 | 60 |
| sq2s | 600 | 1.1E-02 | 1.1E+04 | 121 |
| sq2s | 1000 | 5.3E-03 | 3.8E+04 | 202 |
| e2s | 400 | 2.2E-04 | 1.1E+06 | 245 |
| e2s | 800 | 5.2E-05 | 6.5E+06 | 342 |
| e2s | 1200 | 2.8E-05 | 1.5E+07 | 417 |
| s2s | 200 | 4.0E-02 | 4.8E+03 | 193 |
| s2s | 400 | 1.9E-02 | 1.1E+04 | 207 |

Table 2 Regression coefficients derived from 2007 data.

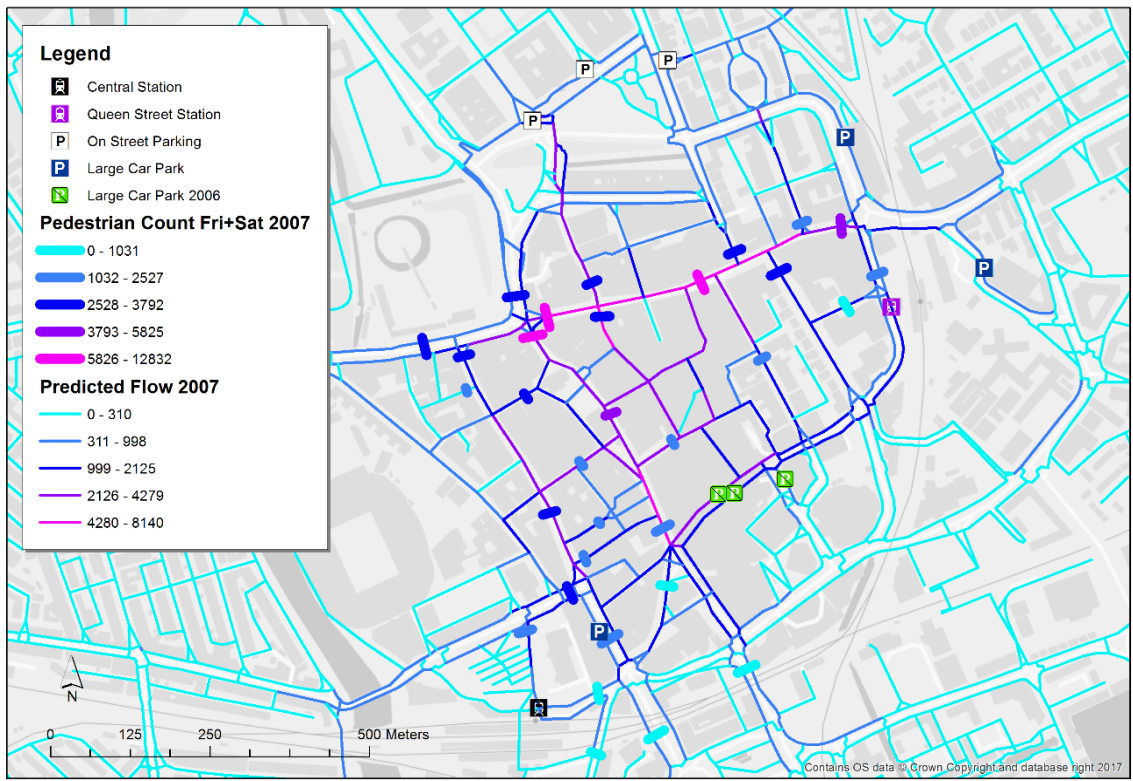

Figure 5 Extrapolation across space: predicted flows for the 2007 map calibrated to 2007 counts

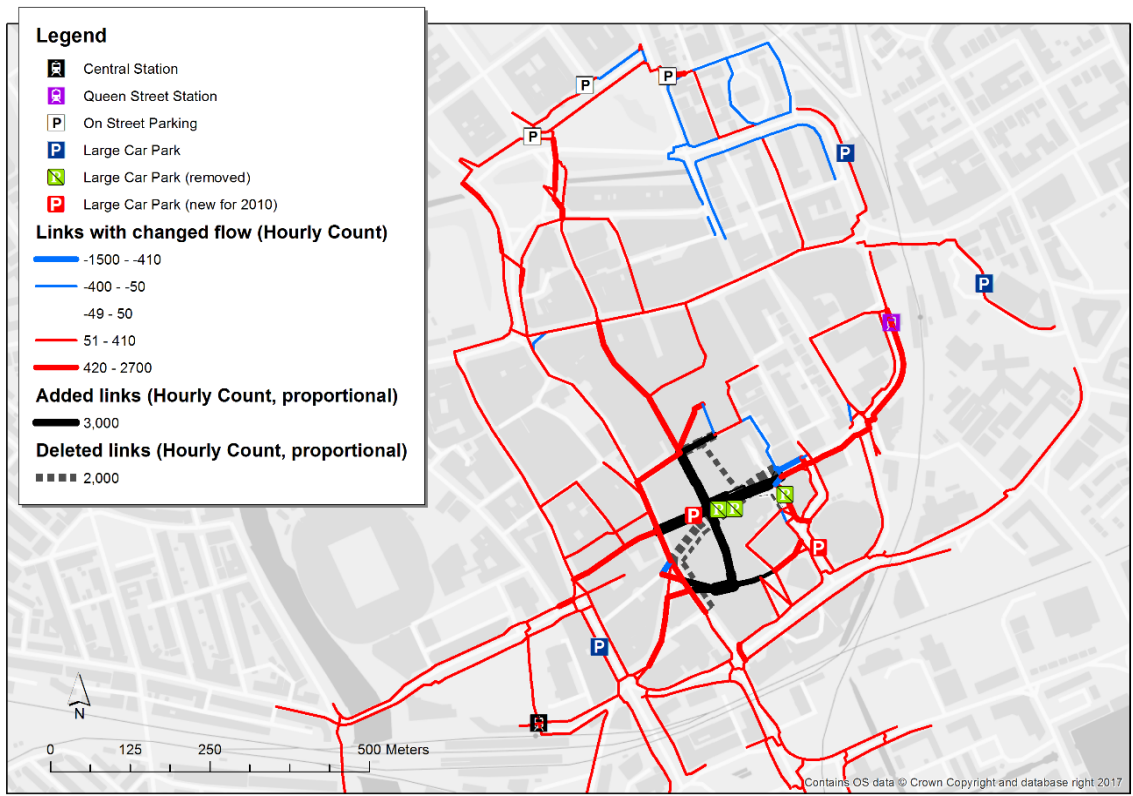

*Figure 6 Extrapolation across time: predicted changes for the 2010 map based on calibration from 2007*

Figure 5 shows pedestrian flows fitted to the footfall count data for 2007. This reveals progressively higher pedestrian flows from the edge of the city to the centre with the highest flows occurring in and around the pedestrianised streets, as would be expected. It is also clear how side streets off busy streets can have much lower pedestrian flows despite linking streets with high levels of flows. The carparks and the Central station do not appear to influence larger flows, probably due to the latter being quite evenly distributed around the periphery.

Figure 6 shows the predicted change in pedestrian flows between 2007 and 2011. The thick red lines shows the greatest increase in flows with blue showing a decrease; white reveals no substantial change. What this clearly predicts is the increase in flows to the new St David's 2 development in the bottom centre of the map and in particular the importance of the streets linking the new development to the two railway stations and bus station and also the main pedestrianised thoroughfare from the north of the city centre. The smaller side streets show no predicted change in flows except where close enough to the new development (400m) that they can capture 'shop-to-shop' traffic with their own retail area. The area to the north-east of the city centre sees slight predicted decline due to inelastic shop-to-shop flows originating on Queen Street, being redistributed from this area to the new development. Note that as total volume of pedestrian activity as exogenous to the model, this predicted decline is relative to the rest of the city centre and may not represent an absolute decline in pedestrian flows.

Table 3 shows model performance for each year. The null model reveals a substantial consistency for pedestrian flows between years, with the exception of 2011 which was hard to predict for all models including the null model. The incremental model outperforms the null model; the performance of the (most widely applicable) direct model is good at 0.72 in 2010. Figure 7 shows a scatter plot of 2010 counts against predictions with clear correlation; no pattern is immediately discernible in the residuals when mapped.

| Year | Null model r2 | Incremental model r2 | Direct model r2 |
| --- | --- | --- | --- |
| 2008 | 0.79 | n/a | n/a |
| 2009 | 0.85 | n/a | n/a |
| 2010 | 0.81 | 0.84 | 0.72 |
| 2011 | 0.63 | 0.73 | 0.45 |

*Table 3 Performance (unweighted r2) of each model in prediction of flows for subsequent years*

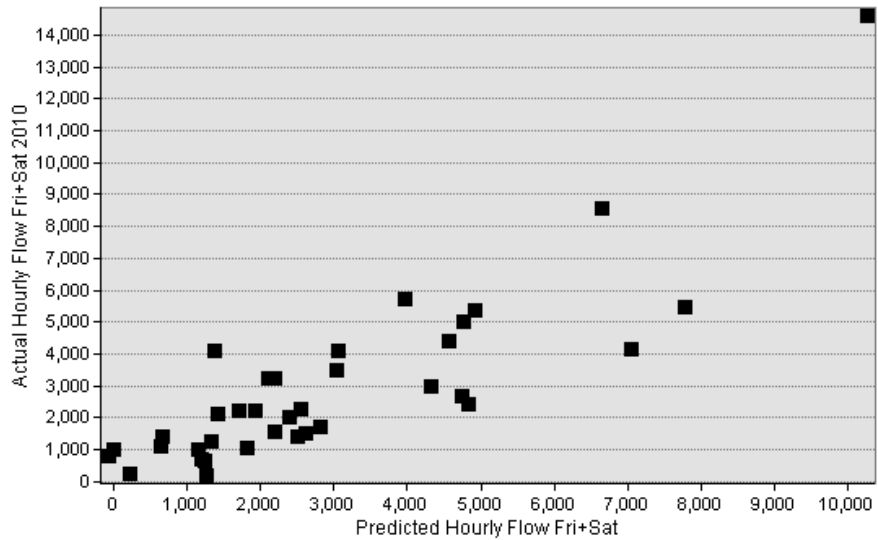

*Figure 7 Scatter plot of 2011 counts against predictions from the direct model calibrated on 2007*

## Discussion

This study has shown Multivariate Hybrid Spatial Network Analysis to be capable predicting changes in vitality related to town centre street layout, including urban block size changes by extrapolating from measured pedestrian flow data both across space and across time. This is the first time, to our knowledge, that a pedestrian flow model has been evaluated for its ability to forecast the effect of town centre changes over time. The methodology used enables better understanding of the impacts of "town centre first" retail-led urban regeneration policy, from the point of view of town centre vitality and pedestrian users rather than retailers alone (Kim and Jang, 2017).

A key limitation of the study is the methodology used for pedestrian counts, which were manual. Modern counting techniques including video analysis, WiFi or Bluetooth sensors can count passing pedestrians 24 hours a day, 365 days a year, with the hope of producing more accurate data. There is also no pedestrian flow measurement from inside the new St Davids 2 shopping centre, which would have been beneficial to include if present.

Two obvious options exist for improving the models outlined here. The first is the inclusion of a pedestrian environment audit; this need not be as complex as mainstream audits (Transport for London 2006) as a simple means of classifying streets e.g. as excellent/good/mediocre/poor - to feed into the route choice model and is likely to yield improvements without excessive cost. The second, as we start to model more congested urban environments, is to account for pedestrian congestion: too much vitality can be seen as the opportunity to expand the town centre or seek alternative sub-centre development; the logical next steps following a policy of "town centre first". In the context of MHSpNA this can be achieved in two ways: either linking to agent microsimulation models at key congested locations, or by iterative modelling using a statistical physics approach (e.g. Osaragi 2004) that predicts deterrence from links based on their width and current level of pedestrian flow.

Finally it would be fruitful to employ MHSpNA techniques to improve the accuracy of existing mode choice models (covering the decision to walk rather than drive e.g. Ewing et al. 2014), thus expanding the social/economic sustainability concern of town centre vitality to incorporate the environmental as well.

## Appendix 1: Model Fitting

Statistical models are fitted as per Cooper (2018). Spatial Network Analyses are typically univariate, that is, they involve only one betweenness calculation, which is then calibrated against vitality as pedestrian flows through bivariate ordinary least squares linear regression. Model fit against data is reported, but not validated against a test data set. Both betweenness and flow variables are often transformed prior to regression, e.g. by cube root (e.g. Turner 2007) or Box Cox estimation (e.g. Cooper 2015). This serves the dual purpose both of taming outliers in the data, and minimizing a trade-off of absolute and relative error.

These techniques are not suitable for a multivariate analysis, because (1) variable transformations violate the structure of what should be a linear additive model, and (2) ordinary least squares will tend to overfit data where the predictor variables are correlated, which is almost always the case in multiple betweenness calculations on the same network. We therefore replace variable transformations with weighting to achieve a similar effect, at the expense of some loss in model fit, but producing more credible models as they are structurally and behaviourally sound. Each data point y is weighted by $y^\lambda/y$, with $0 \leq \lambda \leq 1$ (λ is set to 1 to minimize absolute error, 0 to minimize relative error, or any value in between for a trade-off).

In place of ordinary least squares regression we use Tikhonov regularization in the form of ridge regression (Amemiya 1985; Tikhonov 1943). This technique can be interpreted either as introducing a penalty term to prevent overfit, or as imposing a Bayesian prior on the likely values of the regression coefficients, forcing them towards zero. The optimum strength of the ridge penalty (or standard deviation of the prior) is determined by generalized cross-validation (GCV) with 7 folds and 50 bootstrap repetitions. GCV repeatedly fits models on a random subset of the training data, then tests them on the remainder. This not only solves the problem of overfit, but also has the result of reporting the model's ability to predict outside the training set, i.e. to extrapolate from count points in the training set to the rest of the network. For the ridge regression it was necessary to manually specify the ridge penalty λ as autoselection of λ from glmnet did not find the optimal value.

## Acknowledgements

We thank Martin Wedderburn for helpful discussions, and Antoine Chanthavisouk for his initial work on the 2011 network.

This study was supported by a Cardiff University Data Innovation Research Institute grant.

Network based on Ordnance Survey ITN and Mastermap © Crown Copyright and Database Right 2017 Ordnance Survey (Digimap Licence). Basemaps are Ordnance Survey Open Grey © Crown copyright and database right (2017).